\documentclass[twocolumn]{aastex62}
\pdfoutput=0 
\usepackage{amsmath,amstext}
\usepackage[T1]{fontenc}
\usepackage{apjfonts} 
\usepackage{graphicx}
\usepackage[caption=false]{subfig}
\usepackage[figure,figure*]{hypcap}

\def\geqsim{\lower.73ex\hbox{$\sim$}\llap{\raise.4ex\hbox{$>$}}$\,$}
\def\leqsim{\lower.73ex\hbox{$\sim$}\llap{\raise.4ex\hbox{$<$}}$\,$}

\received{December 19, 2018}
\revised{April 11, 2019}
\accepted{April 15, 2019}
\submitjournal{The Astrophysical Journal}

\shorttitle{Radial Velocity Variability in Dwarf Carbon Stars}
\shortauthors{Roulston et al.}

\turnoffediting

\begin{document}

\title{The Time-Domain Spectroscopic Survey: Radial Velocity Variability in Dwarf Carbon Stars}

\correspondingauthor{Benjamin Roulston}
\email{roulstbr@bu.edu}

\author[0000-0002-9453-7735]{Benjamin R. Roulston}
\altaffiliation{SAO Predoctoral Fellow}

\affiliation{Department of Astronomy, Boston University, 725 Commonwealth Avenue, Boston, MA 02215, USA}
\affiliation{Center for Astrophysics | Harvard \& Smithsonian, 60 Garden St, Cambridge, MA 02138, USA}

\author[0000-0002-8179-9445]{Paul J. Green}
\affiliation{Center for Astrophysics | Harvard \& Smithsonian, 60 Garden St, Cambridge, MA 02138, USA}

\author[0000-0001-8665-5523]{John J. Ruan}
\affiliation{McGill University, 3550 University Street Montreal, QC H3A 2A7, Canada}

\author{Chelsea L. MacLeod}
\affiliation{Center for Astrophysics | Harvard \& Smithsonian, 60 Garden St, Cambridge, MA 02138, USA}

\author{Scott F. Anderson}
\affiliation{Department of Astronomy, University of Washington, Box 351580, Seattle, WA 98195, USA}

\author[0000-0003-3494-343X]{Carles Badenes}
\affiliation{Department of Physics and Astronomy and Pittsburgh Particle Physics, Astrophysics and Cosmology Center (PITT PACC), University of Pittsburgh, 3941 O'Hara St, Pittsburgh, PA 15260, USA}
\author[0000-0002-8725-1069]{Joel R. Brownstein}
\affiliation{Department of Physics and Astronomy, University of Utah, 115 S. 1400 E., Salt Lake City, UT 84112, USA}

\author{Donald P. Schneider}
\affiliation{Department of Astronomy and Astrophysics, The Pennsylvania State University, 525 Davey Laboratory, University Park, PA 16802, USA} 

\author[0000-0002-3481-9052]{Keivan G. Stassun}
\affiliation{Department of Physics \& Astronomy, Vanderbilt University,  6301 Stevenson Center Lane, Nashville, TN 37235, USA} 

\begin{abstract}
\added{\edit1{Dwarf carbon (dC) stars, main sequence stars showing carbon molecular bands, were initially thought to be an oxymoron since only AGB stars dredge carbon into their atmospheres. Mass transfer from a former AGB companion that has since faded to a white dwarf seems the most likely explanation.  Indeed, a few types of giants known to show anomalous abundances --- notably, the CH, Ba and CEMP-s stars --- are known to have a high binary frequency.  The dC stars may be the enhanced-abundance progenitors of most, if not all, of these systems, but this requires demonstrating a high binary frequency for dCs. Here, for a sample of 240 dC stars targeted for repeat spectroscopy by the SDSS-IV's Time Domain Spectroscopic Survey, we analyze radial velocity variability to constrain the binary frequency and orbital properties.  A handful of dC systems show large velocity variability ($>$100 km s$^{-1}$). We compare the dCs to a control sample with a similar distribution of magnitude, color, proper motion, and parallax. Using MCMC methods, we use the measured $\Delta$RV distribution to estimate the binary fraction and the separation distribution assuming both a unimodal and bimodal distribution. We find the dC stars have an enhanced binary fraction of 95\%, consistent with them being products of mass transfer. These models result in mean separations of less than 1 AU corresponding to periods on the order of 1 year. Our results support the conclusion that dC stars form from close binary systems via mass transfer.
}}
\end{abstract}
 

\section{Introduction}
The first carbon star was discovered by \citet{secchi1869} showing strong bandheads of C$_{\rm 2}$ in their optical spectra.  The phrase ``dwarf carbon star'' would have long been considered an oxymoron since it was assumed that all carbon stars were giants AGB stars that had dredged up carbon produced in their cores which polluted their atmospheres, producing an atmospheric C/O ratio above unity.  Carbon preferentially binds with oxygen to form CO. Remaining carbon can combine to form the carbon compounds C$_2$, CN, and CH giving rise to strong carbon molecular bandheads dominating their optical spectra.

This assumption was shown to be invalid when \citet{dahn77} discovered the first main-sequence carbon star, the dwarf carbon (dC) star G77-61.  How could a main-sequence star have enough carbon in its atmosphere to push the C/O ratio beyond unity and create the strong carbon bandheads seen? The currently favored hypothesis is that dC stars are the products of binary mass transfer where the former AGB companion has become a white dwarf, leaving a carbon-enhanced dC primary.   Indeed, about a dozen ``smoking gun'' systems, having composite spectra with a hot DA white dwarf component \citep{hebert93,liebert94,green13,si14},  bolster this hypothesis.  \added{\edit1{There is mounting evidence that many dC stars belong to a metal-poor halo population \citep{farihi18}, wherein lower metallicity may reduce the amount of mass transfer required to achieve $C/O>1$.}}

As main sequence stars with carbon-enriched atmospheres, dC stars are probably the progenitors of the typically more luminous carbon-enhanced metal-poor (CEMP), sgCH, CH, and, perhaps, barium (Ba II) stars, which all have carbon and $s$-process enhancements (see discussion and references in \citealt{Jorissen2016, demarco17}). Samples of such stars have all been targets of spectroscopic radial velocity (RV) monitoring campaigns to test for binarity and characterize orbits. \citet{Sperauskas16} recently studied the CH-like stars --- giants showing CH and $s$-process abundances like CH stars, but without their halo kinematics --- and showed $\sim6$ times higher RV variability for C-rich than C-normal giants.
\citet{Leiner} and others reported that binary interactions occur with surprising frequency, modifying the evolution of a considerable fraction of stars.
Binary systems with mass transfer follow an impressive variety of evolutionary channels and may result in important and spectacular systems such as luminous red novae, Type\,Ia supernovae, planetary nebulae, and more.  Here, we demonstrate that dCs definitively belong to the family of mass transfer binary systems.

The first dC star confirmed to be in a binary system, with a measured period of 245d, is the dC prototype G77-61 \citep{dearborn86}.  \citet{margon18} recently mapped the orbit of a second system with a much shorter (3d) period. \added{\edit2{\citet{Harris2018} found 3 dC stars to be astrometric binaries with periods of 1.23, 3.21, and 11.35 yr (the amplitudes of the photocenter orbits of these three are 0.9, 1.6, and 3.1 au respectively).}} \citet{whitehouse18} detected radial velocity variability in 21 of 28 dC stars monitored spectroscopically with the William Herschel Telescope between 2013 and 2017.  

We seek here, using multi-epoch spectroscopy of many dC stars discovered by the Sloan Digital Sky Survey \citep[SDSS;][]{blanton17}, to measure the dC binary frequency, which should be near unity in this mass transfer scenario. As was the case in \citet{whitehouse18}, our SDSS sampling lacks enough epochs to determine individual orbital parameters.  However, with a significantly larger sample of 240 dC stars, we can use the distribution of radial velocity variations and Markov chain Monte Carlo methods to characterize the dC population's \added{\edit1{binary fraction and the}} separation distribution as was done by \citet{maoz12} and \citet{Badenes2012}.  

\section{Dwarf Carbon Star Sample Selection}
\label{dC_sample_selection}

Dwarf carbon stars for this study were selected from the \citet{green13} and \citet{si14} carbon star samples.  \citet{green13} identified carbon stars by visual inspection of single-epoch SDSS spectra compiled from the union of (1) SDSS DR7 spectra \citep{DR7} having strong cross-correlation coefficients with the SDSS carbon star templates with (2) SDSS spectra with a DR8 pipeline class of STAR and subclass including the word carbon \citep{DR8}.  
From within this carbon star parent sample, definitive main sequence dwarfs were selected by having significant proper motions ($\geq 3\sigma$ and 11\,mas yr$^{-1}$) in the catalog of \citet{munn04} and/or having SDSS spectra visibly identifiable as composite DA/dC spectroscopic binaries \added{\edit1{(there are 3 DA/dC composites in our sample)}}. 
\citet{si14} selected dCs using \added{\edit1{a label propagation algorithm from SDSS DR8, yielding 96 new dC stars.}}

For our current work, which aims to measure radial velocity variability, the primary additional selection criterion was that the selected dC stars have more than one epoch of spectroscopy in the SDSS as of November 2017.   The majority of such objects were intentionally targeted for a second epoch of spectroscopy with the Time Domain Spectroscopic Survey \citep[TDSS;][]{tdss}, a subprogram of the SDSS-IV  extended Baryon Acoustic Oscillation Sky Survey  \citep[eBOSS;][]{dawson16} project.  Within TDSS, the main single-epoch-spectroscopy (SES) program \citep{morganson15} --- along with its  pilot survey, dubbed SEQUELS within SDSS-III \citep{ruan16} --- primarily targets optical point sources (unconfirmed quasars and stars) for a first epoch of spectroscopy based on variability.  However, within several ``few-epoch spectroscopy'' (FES) subprograms, TDSS also acquires repeat spectroscopic observations for subsets of known stars and quasars that are astrophysically interesting.  The FES programs are described by \citet{macleod18} and include several classes of quasars and stars, including dC stars, re-targeted to study their spectroscopic variability. For the dC FES program, we selected all 730 SDSS dC stars from \citet{green13} as well as another 99 dC stars found by \citet{si14}, totaling 829 unique dC stars \added{\edit1{provided as candidates for spectroscopy by the SDSS-IV eBOSS project.}}  \added{\edit1{About 40\% of those stars are expected to be observed by the end of the eBOSS survey.}}

\added{\edit1{The observations for this work (for both dC and control sample) are from a combination of SDSS-I/SDSS-II and SDSS-III/SDSS-IV spectroscopic data. SDSS-I/SDSS-II spectra were taken with the legacy SDSS spectrograph. These data have a wavelength range of 3900--9100\AA\ with a resolution of $R \sim 2000$. The pixel size is 69 km s$^{-1}$. The new eBOSS spectrograph \citep{eboss_spec} in SDSS-III/SDSS-IV has improved qualities. This spectrograph covers the 3,600--10,400\AA\ range and has a resolution of $R \sim 2500$. This spectrograph has a 1.7\AA\ per pixel size.}}

We then searched the SDSS database (using the CasJobs query tool) for spectroscopy from DR14 \citep{dr14} for the dC stars that have been observed in the dC FES program. We also checked the DR14 database for all dC stars in the \citet{green13} sample in search of any dC stars that may have been observed more than once, but not as part of the TDSS FES program. Our final sample contains FES spectra obtained up until October 31, 2017 and spectra from DR14. 

We visually inspected all spectral epochs and removed any spectra that had strong broad artifacts. The final sample for this study was 240 dCs with a total of 540 spectra within the SDSS. 

\begin{figure} 
\centering
\epsscale{1.2}
\plotone{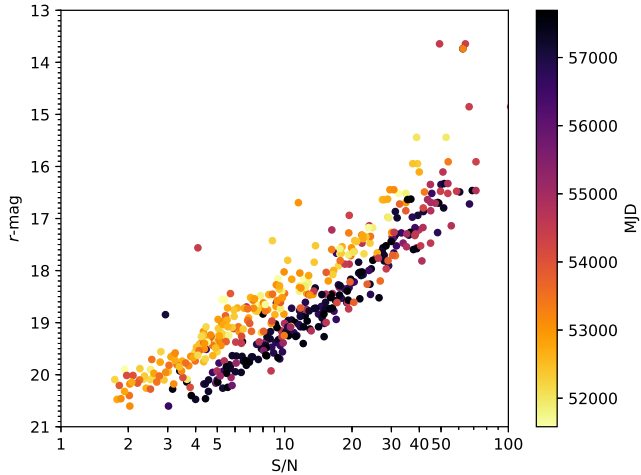}
\caption{Optical $r$-band magnitude as a function of the median spectroscopic signal-to-noise ratio for our sample. Dots are colored by spectroscopic epoch in modified Julian days (MJD). The improved $S/N$ of later epochs (using the BOSS rather than the legacy SDSS spectrograph) is evident.  As expected, there is a close correlation between brightness and $S/N$.  }
\label{fig:rmag_snmed}
\end{figure}

Figure \ref{fig:rmag_snmed} shows the correlation between $r$ mag \citep{SDSS_rmag} and the median spectroscopic signal-to-noise ratio (S/N). The color axis is the Modified Julian Date (MJD) of each epoch, and a clear distinction can be seen between the early epochs and later epochs in regards to S/N due to the enhanced capabilities of the BOSS spectrograph.

\section{Control Sample Selection}
\label{control_sample_selection}

\subsection{Selection Criteria}
\label{control_selection_criteria}
The control sample was selected from the SDSS DR14 catalog using the properties of the dC sample as a selection criteria. The control sample criteria were as follows: (1) objects must have CLASS=STAR from the  SDSS spectroscopic pipeline (2) a significant proper motion detection following the criteria of \citet{green13}\footnote{Proper motion in at least one coordinate larger than $3\sigma$ where $\sigma$ is the proper motion uncertainty in that coordinate, and total proper motion larger than 11~mas\,yr$^{-1}$.}  (3) select only stars within the 2 -- 98\% parameter ranges of the dC sample (i.e., total proper motion between 11 and 143 milliarcseconds yr$^{-1}$, SDSS $r$ mag between 15.9 and 20.3, and a $g-r$ color between 0.375 and 1.908 using extinction-corrected magnitudes and colors). All carbon stars (including dCs) were removed from the sample by SDSS CLASS and SUBCLASS keywords and by matching to all known dCs.  Since their binary fraction is likely to be highly biased, we further removed stars originally targeted for reasons of X-ray emission or variability.\footnote{We removed from the control sample any eBOSS\_TARGET0 stars that are selected by variability as TDSS target (8). Most of these variables are RR Lyr or close eclipsing binaries and some are dC stars. We further removed stars where LEGACY PrimTarget keyword contained ROSAT or where BOSS ANCILLARY\_TARGET1 = QSO\_VAR, QSO\_VAR\_LF or QSO\_VAR\_SDSS. Finally, common proper motion binaries were removed by eliminating control stars with BOSS ANCILLARY\_TARGET2 = SPOKE2.} Finally, all control stars were required to have a match in the Gaia DR2 data release \citep{gaia2}.  These criteria returned 9,822 stars that had more than one SDSS spectrum for a total of 21,820 spectra.

\subsection{Property Matching}
\label{control_sample_distance}
To reduce the effects of differing properties between the dC sample and the control sample, we matched the control stars to each dC by finding the normalized ``distance'' in a ``four-property space'': $r$ mag, $g-r$ color, Gaia DR2 total proper motion, and Gaia DR2 parallax.\footnote{We use parallax rather than  distance due to the subtleties of converting Gaia DR2 parallaxes to distance as noted in \citet{gaia_parallax}} 

This distance matching was performed by creating a ``normalized coordinate'' out of each of the four properties. This coordinate was constructed by subtracting the minimum property value, then dividing by the maximum value for the property. This approach scales all of the values for each property into the range of $[0,1]$ based on the dC sample 
so that all of the properties are similarly weighted. 

\begin{figure} 
\centering
\epsscale{1.2}
\plotone{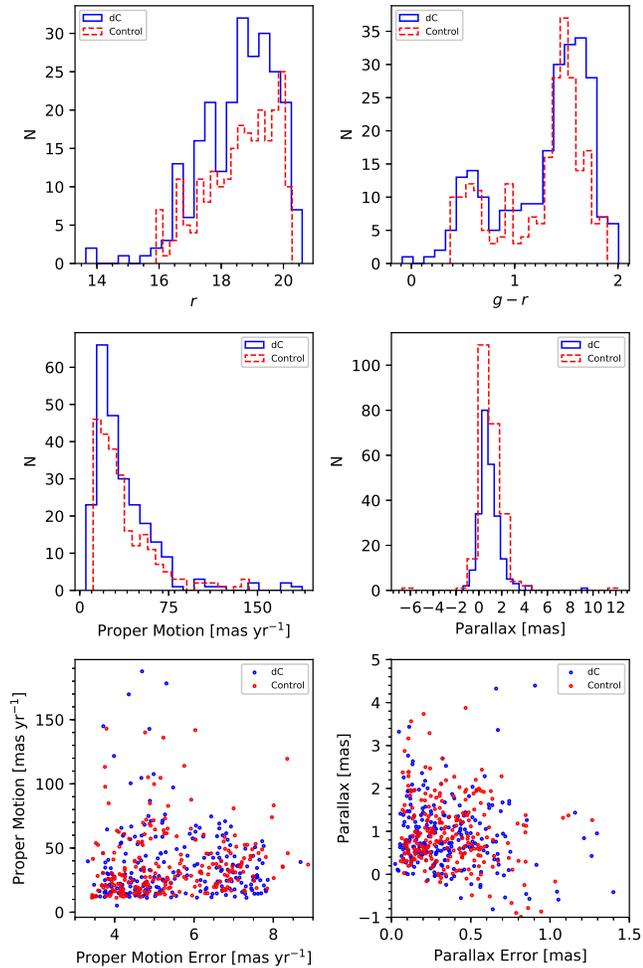}
\caption{Comparison plot for the four properties used to match the control sample to the dC sample. The first four panels are histograms of the four properties used in the control sample matching process. The dC sample is in solid blue lines, and the control sample is in dashed red lines. Our matching process is designed to recreate the dC histograms with the control sample. \added{\edit2{The bottom two panels are scatter plots of total proper motion and parallax with their associated errors. From these it is clear that the control sample has similar errors in proper motion and parallax as the dC sample, although not matched on those errors.}}}
\label{fig:sample_histos}
\end{figure}

These coordinates were used to find the distance from each dC to all of the control stars. These distances, once sorted, give the closest matching control stars for each individual dC based on the chosen four properties. With the control sample sorted for closest matching properties to the dC sample, we drew the closest matches for each dC to create the final, property-matched control sample to analyze along the dC sample. Figure \ref{fig:sample_histos}, compares histograms of these four properties \added{\edit2{(and errors on proper motion and parallax)}} for the dC and control samples. 

\subsection{Control Sample Issues}
\label{control_sample_issues}
The control sample, even given the matching process we used, is not perfect for several reasons.

(1) The SDSS stellar sample \added{\edit1{was}} produced by a hodgepodge of different targeting programs, some of which may skew the $\Delta$RV distribution.  

(2) It could be more difficult to detect binarity in the control sample because the single spectrum of an unresolved binary contains (by definition) both components. If the two components have significantly different main sequence spectral types or evolutionary stage (e.g., giant + dwarf), then one component is much more luminous than the other --- similar to the dC systems we expect, which likely contain a white dwarf too cool to detect in most spectra.  However, if the two components have close spectral types (e.g., a K7+M2 binary), they contribute similarly to the spectral flux. Thus, the observed velocity changes are muted because if one component is approaching, the other is receding at any epoch.  There are techniques that could mitigate this issue such as attempting to fit the sum of two spectral templates to each spectrum (e.g., as proposed by \citealt{el-badry18}, but this approach would be effective only for some combinations of mass ratios and S/N.

(3) The control sample has a significantly different MJD sampling than that of the dC sample. A majority of the control sample was observed in the earliest versions of the SDSS and have $\Delta$MJDs between spectroscopic epochs on average of 100 days. Most of the dC stars have been specifically targeted by TDSS for repeat spectroscopy during SDSS-IV; so, they have a $\Delta$MJD distribution of typically 1000s of days. While this sampling does affect the range of periods our methods are sensitive to, it should not impact our results. Since we are searching for close binary systems which have large $\Delta$RVs and, therefore, short periods, the control sample's accessible  $\Delta$MJD distribution would only impede our ability to detect wide binary systems for which our sensitivity is already severely limited by the RV errors \added{\edit1{as shown in Section \ref{sec:xc_err}}}. 

The first two items are observational and may diminish the discriminating power of our tests.  Other intrinsic differences may complicate our analysis and interpretation of the results.  For instance, we expect dC stars to have a 100\% binary fraction, but a very narrow range of companion masses (all white dwarfs, therefore, strongly peaked near 0.5$\,M_{\odot}$).  By contrast, the control sample has a certain binary fraction, but the distribution of companion masses in those binaries will have a wider range.  The orbital properties of binaries in the control sample may also have a wider range.  We expect that the dC has interacted with its (former AGB) companion (e.g., either by wind accretion or Roche lobe overflow) which sets upper (and perhaps even lower) limits on the orbital separation.  The only effective limit on orbital separation in the control sample is that the pair be spatially unresolved ($\leqsim\,2\arcsec$).

\section{Radial Velocity Analysis}
\label{RVs}

\subsection{Cross-Correlation Method}
\label{CrosscorrMethod}

We measured radial velocity variations ($\Delta$RV)  using the \texttt{IRAF}\footnote{IRAF is distributed by the National Optical Astronomy Observatories,which are operated by the Association of Universities for Research in Astronomy, Inc., under cooperative agreement with the National Science Foundation.} \citep{iraf} package \texttt{FXCOR} that cross-correlates between a template and object spectrum following the methods of \citet{tonry79}.

Each spectrum was visually inspected to insure the S/N was sufficiently high for cross-correlation as well as to identify wavelength regions with corrupt data. We also searched for any problematic features that could affect the cross-correlation. Those objects that had corrupted regions were marked and individually run through the cross-correlation, ignoring those corrupted regions. The rest of the sample was cross-correlated in a batch, all using the same constraints and regions.

Each epoch combination's cross-correlation function was manually inspected to check the quality of the cross-correlation.  In a small ($\sim$10\%) fraction of cases, the cross-correlation function is best fit manually. If the cross-correlation function could not be fit (e.g., no peak in the CCF is preferred), that epoch combination was removed from the sample.

\subsection{Cross-Correlation Errors}
\label{sec:xc_err}
To validate the cross-correlation process, we ran a variety of tests. The first was to verify and, if possible, minimize the reported errors from \texttt{FXCOR}.

To minimize uncertainties in the $\Delta$RV measurements given by the cross-correlation, we \added{\edit1{used}} two techniques: (1) direct cross-correlation of one object against itself across different  epochs and (2) cross-correlation of each epoch for one object against a SDSS C star template spectrum. For each method we also experimented with changing the regions sampled (e.g., only narrow atomic lines, excluding the carbon bands, or only including carbon bands).

\replaced{From all combinations, we adopted, as the best method, the direct cross-correlation between two epochs for a single object using the spectrum in the range of $4000$ \AA$ - 7000$ \AA, ignoring telluric line regions.}{\edit1{From all combinations, we found the best method to be the direct cross-correlation between two epochs for a single object using the spectrum in the range of $4000$ \AA$ - 7000$ \AA, ignoring telluric line regions, which is the method we adopt for this work.}}  We use one epoch (the early MJD) as the ``template'' and the other epoch (the later MJD) as the ``object''. This method produces some benefits over using the usual template method: (1) This cross-correlation directly provides the $\Delta$RV shift. \replaced{(2) Since we use the same dC as the template and the object spectra, which reduces the $\Delta$RV errors.}{\edit1{(2) Since we use the same dC as the template and the object spectra, the $\Delta$RV errors are reduced because a star is its own perfect template.}} (3) The SDSS C star template spectrum is for AGB C stars; there are no templates for dC stars. 

The second test performed was to determine if the reported values and errors from \texttt{FXCOR} are believable for both dC and control spectra. \replaced{This test involved measuring the $\Delta$RV for each object between two epochs to obtain the baseline shift. We then artificially shifted the later epoch by 30 different shifts between -100 km s$^{-1}$ and 100 km s$^{-1}$, then reran the cross-correlation to see how accurately \texttt{FXCOR} recovered each of those applied shifts between differing epochs.}{\edit1{This test involved finding ``multi-shift'' errors for our objects by trying to recover applied shifts between different epochs. We did this by shifting the later epoch by 30 different velocities between -100 km s$^{-1}$ and 100 km s$^{-1}$. Then, using the same cross-correlation setup as we used to measure our $\Delta$RVs, we see how well \texttt{FXCOR} was able to recover our applied shift.}}

\begin{figure} 
\captionsetup[subfloat]{farskip=2pt,captionskip=1pt}
\subfloat[]{\includegraphics[width=\columnwidth]{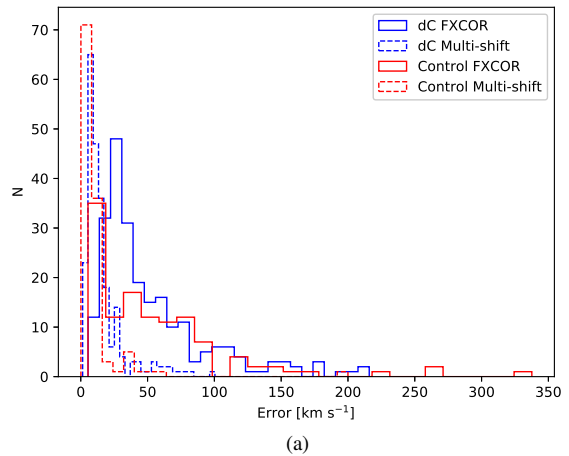}}

\subfloat[]{\includegraphics[width=\columnwidth]{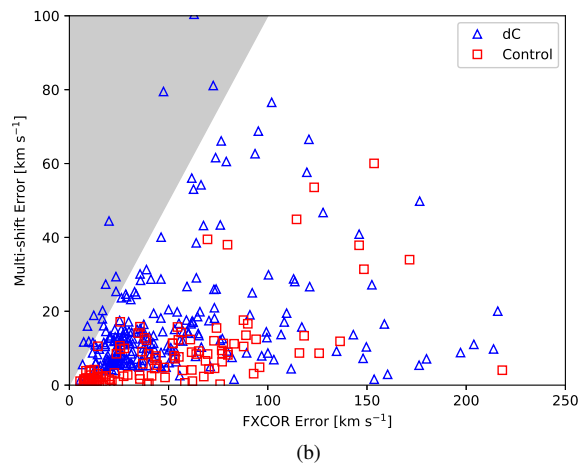}}
\caption{TOP: Histogram of reported \texttt{FXCOR} errors and ``multi-shift'' errors for both the control (red) and dC samples (blue). The errors reported by \texttt{FXCOR} are generally larger than the ``multi-shift'' errors. BOTTOM:  Multi-shift vs. \texttt{FXCOR}-reported errors further show a poor correlation. The shaded region shows those objects where the \texttt{FXCOR}-reported errors are smaller; very few objects lie in this region.}
\label{fig:fxcor_multi-shift_hist} 
\end{figure}

\texttt{FXCOR} was generally able to recover the applied shift in both the dC and control samples. However, the reported errors from \texttt{FXCOR} generally are overestimated. By comparing each object's \texttt{FXCOR}-measured shift for each of the 30 different applied shifts, we determined ``multi-shift'' errors for each sample as the RMS of the measured $-$ applied shift (see Figure \ref{fig:fxcor_multi-shift_hist}).  

Figure \ref{fig:fxcor_multi-shift_hist} presents histograms for the reported \texttt{FXCOR} errors and the measured ``multi-shift'' errors for both the dC and control samples. The top panel (a) shows how across the sample, the errors are smaller for the ``multi-shift'' errors than those reported by \texttt{FXCOR}. The bottom panel (b) displays that as \texttt{FXCOR} error increases, so do the ``multi-shift'' errors (a plausible result as spectral S/N is a key factor in the error).

\added{\edit1{\citet{Kleyna2002} also found that \texttt{FXCOR} errors are overestimated. In their paper, they applied a multiplicative constant re-scaling factor of 0.35 to the \texttt{FXCOR} errors. Using our ``multi-shift'' errors, we find that our combined dC and control sample have an average scaling factor of 0.32. Our scaling is consistent with the value from \citet{Kleyna2002}.}}

\added{\edit1{However, we do not adopt the ``multi-shift'' errors throughout the rest of our analysis. We use the \texttt{FXCOR} reported errors knowing they are overestimated. This allows us to be conservative with the rest of our findings showing our results do not rely on scaling down our errors.}}

\added{\edit1{Assuming $1.0M_\odot$ + WD, edge-on system, given mean $\Delta$RV errors of $\approx 28$ km s$^{-1}$, the longest period we are sensitive to is $\sim 15$ yr. This is much longer then all of the control $\Delta$MJD distribution, and longer than most of the dC $\Delta$MJD distribution, so our sampling is not the limiting factor, the errors are (which is what we would expect). We also control for this by including the $\Delta$MJD distribution in the modeling of Section \ref{sec:orbit_sim} for both the control and dC samples. This involves using the $\Delta$MJD distribution to sample our modeled observations so they represent real observations.}}

\begin{figure} 
\centering
\epsscale{1.2}
\plotone{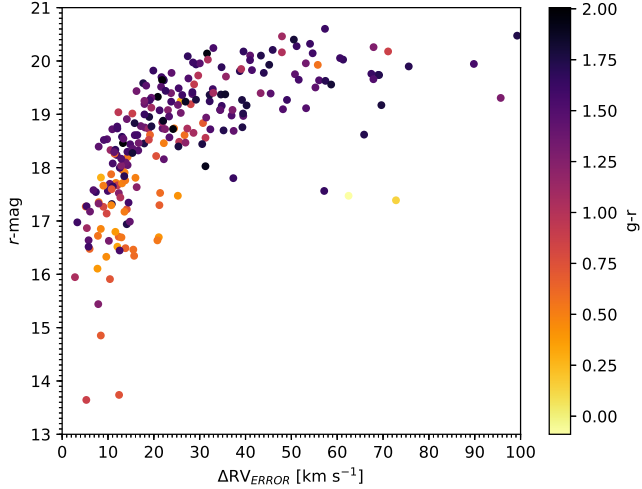}
\caption{Optical $r$-band magnitude as a function of the radial velocity variation  ($\Delta$RV) errors obtained when directly comparing two epochs of dC stars in our sample. Fainter stars have larger errors, as expected, since these tend to have poor spectroscopic $S/N$ (see Fig\ref{fig:rmag_snmed}). Optical $g-r$ color is denoted for each object by color.}
\label{fig:rmag_dRVerr}
\end{figure}

Figure \ref{fig:rmag_dRVerr} shows the relationship between the brightness ($r$ mag) and the $\Delta$RV errors from the cross-correlation. Brightness (and by proxy S/N) determines the $\Delta$RV errors, and bluer objects tend to have smaller errors (again because these stars are usually brighter and have better S/N). 

\subsection{dC and Control $\Delta$RVs}
\label{dRVs}

The dC and control samples were both cross-correlated using the same method. For every object, each possible combination of epochs was cross-correlated (with the earlier epoch as the template and later epoch as the object). From all possible combinations for an object, we selected the maximum $\Delta$RV for our statistical analysis. \added{\edit1{Our samples consist mostly of objects that have only two epochs of spectroscopy. Only a handful (N $\sim 30$) of objects in either sample have more than two epochs.}}

Extremely large $\Delta$RV values (e.g., $>$600 km s$^{-1}$) in a binary with a main sequence component are suspect as in such cases we would expect extremely close orbits and strong signs of interactions and mass transfer. Therefore, any object whose $\Delta$RV was measured to be larger than this value was manually cross-correlated again and had its cross-correlation function manually fit to try and obtain a better $\Delta$RV. If the cross-correlation is unsuitable for fitting, the object was removed from the sample (this only resulted in the removal of two dCs and three control stars). 

\begin{figure}
\centering
\epsscale{1.0}
\plotone{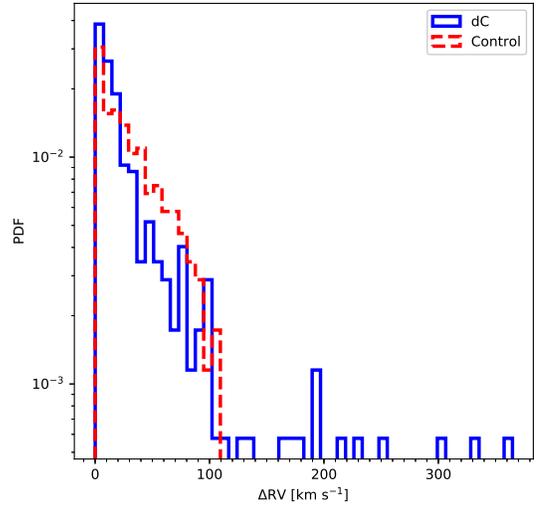}
\caption{Normalized $\Delta$RV histogram for both of the finalized dC and control samples. The wider flaring of the base for the dC sample suggests that dCs are in close binaries.}
\label{fig:dC_dRV}
\end{figure}

Figure \ref{fig:dC_dRV} is a normalized (note the log scale) histogram showing the $\Delta$RV measurements for both the dC (blue) and control (red) samples. The bins are used for each of the samples. This figure demonstrates that both samples have a central core whose width is dominated by the errors. The dC sample, however, has a tail of high $\Delta$RV systems that extends beyond this core. These systems likely represent close binary systems.

In the dC sample, \added{\edit1{we define }}these high $\Delta$RV systems \replaced{are}{\edit1{as}} those objects which display $\Delta$RV values $\geq$ 100 km s$^{-1}$. Stars that display such high $\Delta$RV values are indicative of close binary orbits. \added{\edit1{To confirm these large $\Delta$RV systems, which should have visible shifts in their spectra,}} these large $\Delta$RVs were inspected by shifting the later epoch by the measured $\Delta$RV amount and visually checking to determine if features in the spectrum align. 

\edit2{\deleted{Objects with the highest $\Delta$RV values will be prime targets for determining dC orbits. We have been awarded time on the MMT 6.5-meter to begin dedicated follow spectroscopy to investigate these $\Delta$RV values as well as to enable fitting of orbital parameters, as done recently for SDSS\,J125017.90+252427.6, which had a period 3d and RV amplitude of 100\,km s$^{-1}$ \citep{margon18}.}}

\begin{figure} 
\centering
\epsscale{1.2}
\plotone{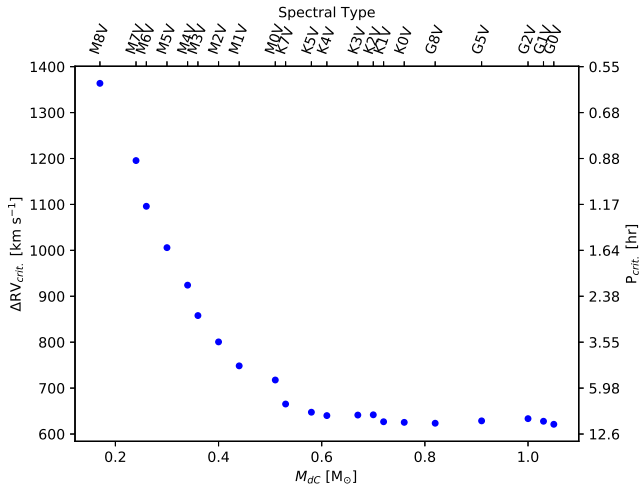}
\caption{The maximum possible value of $\Delta$RV for a range of the expected possible dC masses with a 0.5M$_{\odot}$ WD companion. We define this maximum $\Delta$RV to be when the dC fills its Roche lobe. Using the equation of \citet{Eggleton83}, we calculate the separation for a main-sequence star of each mass to fill its Roche lobe. Also indicated are the corresponding spectral type for each mass and the corresponding period (which is a minimum) for each of the maximum $\Delta$RVs.}
\label{fig:largest_drvs}
\end{figure}

Given that these systems display no strong signs of interaction (such as explosive variability or an accretion disk continuum emission component), few if any of the dCs are likely to have filled their Roche lobes and be transferring mass to the presumed white dwarf companion.  Figure \ref{fig:largest_drvs} shows the largest possible $\Delta$RV we would expect if the dC was filling its Roche lobe (for varying dC masses and a 0.5M$_{\odot}$ WD). Using equation 2 of \citet{Eggleton83}, we calculate the separation for a main-sequence star of each mass to fill its Roche lobe and calculate the corresponding critical $\Delta$RV and period ($\Delta RV_{crit}$,  $P_{crit}$ ) that corresponds to the Roche lobe limit. \replaced{This calculation assumes circular orbits (which we would expect for our dCs) and for the perfect case of edge on systems ($i = 90^\circ$).}{\edit1{This calculation assumes circular orbits (which we would expect for our dCs) and that the dC is the perfect case of an edge on system ($i = 90^\circ$).}} The figure suggests that while we detect dC systems that have large $\Delta$RVs; we have not detected any dCs near the Roche lobe limit \added{\edit1{edge-on}}. 

\begin{figure*} 
\centering
\epsscale{1.2}
\plotone{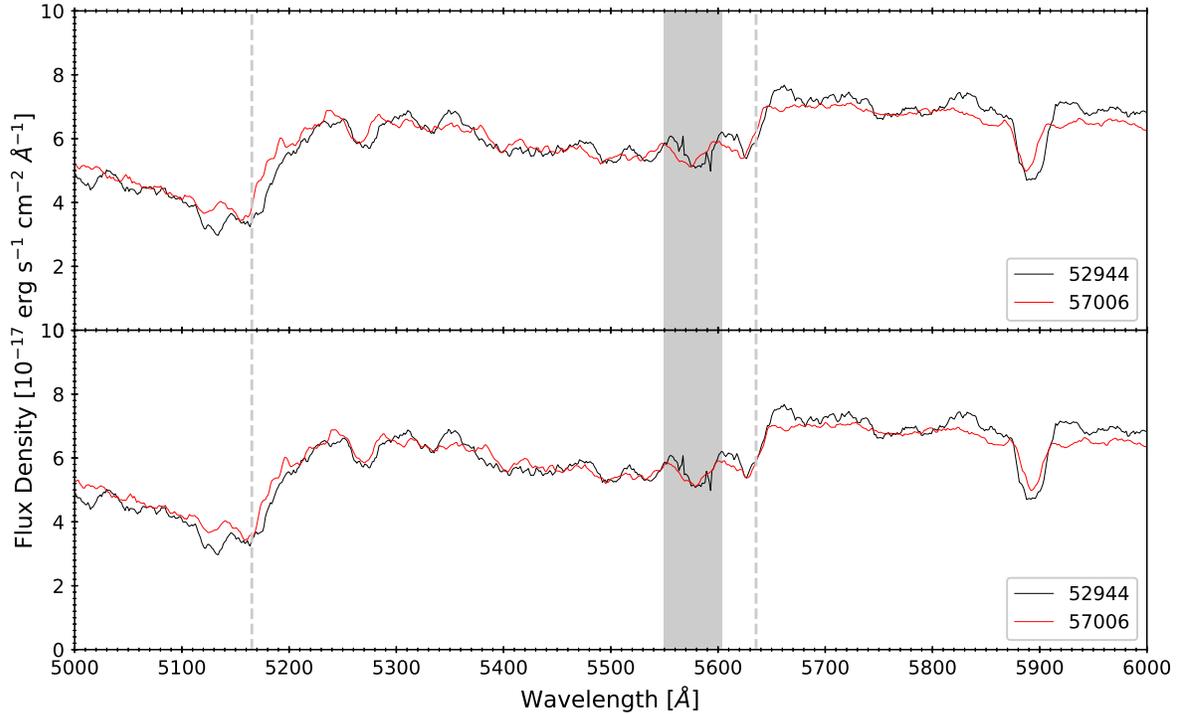}
\caption{Smoothed spectra of both epochs for the dC with one of the largest measured $\Delta$RV, ($\alpha_{J2000}$,$\delta_{J2000}$) = (9.33893$^\circ$, +0.20685$^\circ$). The early epoch is in black, and the later epoch is in red. TOP: shows both epochs as measured in the SDSS. BOTTOM: shows the same spectra, but the late epoch (red) has been shifted by the measured $\Delta$RV = -252 $\pm$ 15 km s$^{-1}$. After shifting, the absorption lines clearly align between epochs confirming this $\Delta$RV and lending evidence that this dC is in a close binary orbit. The gray region is between $5550$ \AA\ and  $5604$ \AA\ and covers the span of the OI night sky line which may contaminate the flux in that region. Locations of the C$_2$ bands in this wavelength range are shown.}
\label{fig:large_believe_dRV}
\end{figure*}

Figure \ref{fig:large_believe_dRV} shows spectra for the dC with one of the largest measured $\Delta$RVs. Both epochs are plotted with the early epoch in black and the late epoch in red. The top panel is of the original spectrum as observed by the SDSS. The bottom panel presents the same epochs, but the later epoch (red) has been shifted by the measured $\Delta$RV = -252 $\pm$ 15 km s$^{-1}$ amount. After this shift, the absorption features in the spectrum align confirming this measured $\Delta$RV. All pre-BOSS spectra in this figure have been smoothed by a box-car of 20 pixels, and all later spectra have been smoothed by a box-car of 15 pixels so the SDSS legacy spectra match the resolution of the new BOSS spectrograph; because otherwise, there is a spurious appearance of variability.

\begin{singlespace}
\begin{deluxetable*}{cDDDDccccccDD}
\tablecaption{dC Sample Properties}
\tablewidth{0.99\textwidth}
\tablehead{\colhead{Index} & \twocolhead{$\alpha_{J2000}$} & \twocolhead{$\delta_{J2000}$} & \twocolhead{$r$} & \twocolhead{$g-r$} & \colhead{Plate1} & \colhead{MJD1} & \colhead{FiberID1} & \colhead{Plate2} & \colhead{MJD2} & \colhead{FiberID2} & \twocolhead{$\Delta$RV} & \twocolhead{$\Delta$RV$_{error}$} \\ \colhead{ } & \twocolhead{[Deg]} & \twocolhead{[Deg]} & \twocolhead{ } & \twocolhead{ } & \colhead{ } & \colhead{ } & \colhead{ } & \colhead{ } & \colhead{ } & \colhead{ } & \twocolhead{[km s$^{-1}$]} & \twocolhead{[km s$^{-1}$]}}
\decimals
\startdata
  1 & 0.0626 & 28.1693 & 17.97 & 0.74 & 2824 & 54452 & 253 & 7696 & 57655 & 83 & 5 & 13 \\
  2 & 0.1483 & -0.1875 & 18.74 & 1.42 & 1489 & 52991 & 156 & 7850 & 56956 & 704 & -18 & 22 \\
  3 & 0.9212 & 23.9270 & 16.94 & 1.74 & 2801 & 54331 & 201 & 7665 & 57328 & 734 & 25 & 14 \\
  4 & 1.2380 & 1.1606 & 18.33 & 1.14 & 1490 & 52994 & 364 & 7862 & 56984 & 626 & 6 & 11 \\
  5 & 3.1909 & -1.0895 & 19.47 & 1.59 & 687 & 52518 & 297 & 7863 & 56975 & 804 & 44 & 33 \\
  6 & 6.8651 & 6.6597 & 17.47 & 0.41 & 3106 & 54714 & 139 & 3106 & 54738 & 131 & 14 & 25 \\
  7 & 7.3310 & 0.7206 & 18.84 & 0.7 & 1087 & 52930 & 504 & 7855 & 57011 & 424 & 26 & 31 \\
  8 & 9.3389 & 0.2069 & 19.01 & 0.88 & 1495 & 52944 & 353 & 7868 & 57006 & 812 & -252 & 15 \\
  9 & 9.9056 & 15.4863 & 18.48 & 0.95 & 419 & 51812 & 346 & 419 & 51868 & 346 & 11 & 26 \\
  10 & 10.8422 & 0.4788 & 16.72 & 0.55 & 1904 & 53682 & 495 & 7870 & 57016 & 562 & 13 & 8 \\
\enddata
\tablecomments{dC property table. Each object can be identified by its SDSS plate-mjd-fiberID combination as well as its celestial coordinates. Included in this table are $r$ mag, $g-r$ color, and the measured $\Delta$RV and errors. This table is sorted on $\alpha_{J2000}$ and each dC is given an index starting at 1. This index links each dC star to the corresponding control star from the matching process. Shown here are the first 10 dC stars. A machine-readable version of the full table is available in the online journal.}
\end{deluxetable*}
\label{tab:dc_prop}

\end{singlespace}
\begin{singlespace}
\begin{deluxetable*}{cDDDDccccccDD}
\tablecaption{Control Sample Properties}
\tablewidth{0.99\textwidth}
\tablehead{\colhead{Index} &\twocolhead{$\alpha_{J2000}$} & \twocolhead{$\delta_{J2000}$} & \twocolhead{$r$} & \twocolhead{$g-r$} & \colhead{Plate1} & \colhead{MJD1} & \colhead{FiberID1} & \colhead{Plate2} & \colhead{MJD2} & \colhead{FiberID2} & \twocolhead{$\Delta$RV} & \twocolhead{$\Delta$RV$_{error}$} \\ \colhead{ } & \twocolhead{[Deg]} & \twocolhead{[Deg]} & \twocolhead{ } & \twocolhead{ } & \colhead{ } & \colhead{ } & \colhead{ } & \colhead{ } & \colhead{ } & \colhead{ } & \twocolhead{[km s$^{-1}$]} & \twocolhead{[km s$^{-1}$]}}
\decimals
\startdata
  1 & 47.6415 & -0.2745 & 17.91 & 0.73 & 2068 & 53386 & 74 & 7255 & 56597 & 160 & 1 & 9 \\
  2 & 172.5454 & 20.1782 & 18.7 & 1.43 & 3170 & 54859 & 640 & 3170 & 54907 & 582 & -54 & 26 \\
  3 & 0.9212 & 23.9269 & 16.94 & 1.74 & 2801 & 54331 & 201 & 7665 & 57328 & 734 & -18 & 19 \\
  4 & 35.3647 & -0.2364 & 18.22 & 1.18 & 704 & 52205 & 234 & 703 & 52209 & 30 & -53 & 19 \\
  5 & 170.6615 & 45.6529 & 19.65 & 1.57 & 3216 & 54853 & 143 & 3216 & 54908 & 156 & 66 & 55 \\
  6 & 6.8651 & 6.6597 & 17.47 & 0.41 & 3106 & 54714 & 139 & 3106 & 54738 & 131 & 22 & 11 \\
  7 & 10.6953 & -0.6110 & 18.83 & 0.7 & 1905 & 53613 & 213 & 1905 & 53706 & 219 & 32 & 27 \\
  8 & 113.8683 & 41.4378 & 19.07 & 0.92 & 3658 & 55205 & 890 & 5941 & 56193 & 884 & 12 & 38 \\
  9 & 44.3177 & 0.9009 & 18.39 & 0.95 & 1512 & 53035 & 590 & 1512 & 53742 & 579 & -2 & 26 \\
  10 & 108.2489 & 38.7804 & 16.77 & 0.54 & 2938 & 54503 & 6 & 2938 & 54526 & 18 & 3 & 8 \\
\enddata
\tablecomments{Control sample property table. Each object can be identified by its SDSS plate-mjd-fiberID combination as well as its celestial coordinates. Included in this table are $r$ mag, $g-r$ color, and the measured $\Delta$RV and errors. This control table is sorted on the index which links each control star to the corresponding dC from the matching process. Shown here are the first 10 control stars that correspond to the first 10 dC stars in Table \ref{tab:dc_prop}.  A machine-readable version of the full table is available in the online journal.}
\end{deluxetable*}
\label{tab:control_prop}

\end{singlespace}

Table \ref{tab:dc_prop} and Table \ref{tab:control_prop} list the properties for the dC and control sample respectively. Only the first 10 rows for each are shown, the full machine readable tables are available online. 

\section{Statistical Comparison of $\Delta$RV Distributions}
\label{dRV_stats}

\subsection{Anderson-Darling Test}
\label{ks_ad}

We used a standard, two-sample Anderson-Darling \citep[AD;][]{ad_test} test to determine the similarity between the dC and control $\Delta$RV distributions. From the measured dC and control $\Delta$RVs, the null hypothesis that the dC and control $\Delta$RVs are drawn from the same distribution can be rejected at the 99.95\% level ($\log{p} = -3.31$).

\subsection{Extreme Deconvolution}
\label{xd}
A drawback of the AD test is that it does not take measurement uncertainties into account when comparing two distributions. For example, two distributions can look dissimilar if their uncertainties are different even if their true underlying distributions are identical. To ensure that the wider $\Delta$RV observed in our dC sample in Figure 5 is not simply due to differences in the measurement uncertainties \added{\edit1{(since the dCs have larger errors, as seen in Figure \ref{fig:fxcor_multi-shift_hist}, likely a result of the C$_2$ and CN bands)}}, we use the extreme deconvolution (XD) method of \cite{xd} to deconvolve the underlying distribution of our $\Delta$RV measurements.
This XD method employs a Gaussian Mixture Model (GMM) to infer the underlying distribution from a set of heterogeneous, noisy observations or samples while incorporating the errors. 

We tested the number of components for the XD-GMM for both the dC and control samples using the Bayesian Information Criterion (BIC) of each model. The BIC approach suggests that the dC sample is best modeled by a mixture of three Gaussians. However, the third Gaussian component for the dC population converges to a small normalization and an unphysically large width; so, we constrain the dC sample to be fit with two components. This decision allows for a central core and for a possible large $\Delta$RV tail that contains close binary systems. \added{\edit1{The control sample is best fit by a single Gaussian as determined by the BIC.}} Table \ref{tab:gmm_parms} lists the parameters for these fit components for both the dC and control XD-GMMs. 

\begin{figure} 
\centering
\epsscale{1.2}
\plotone{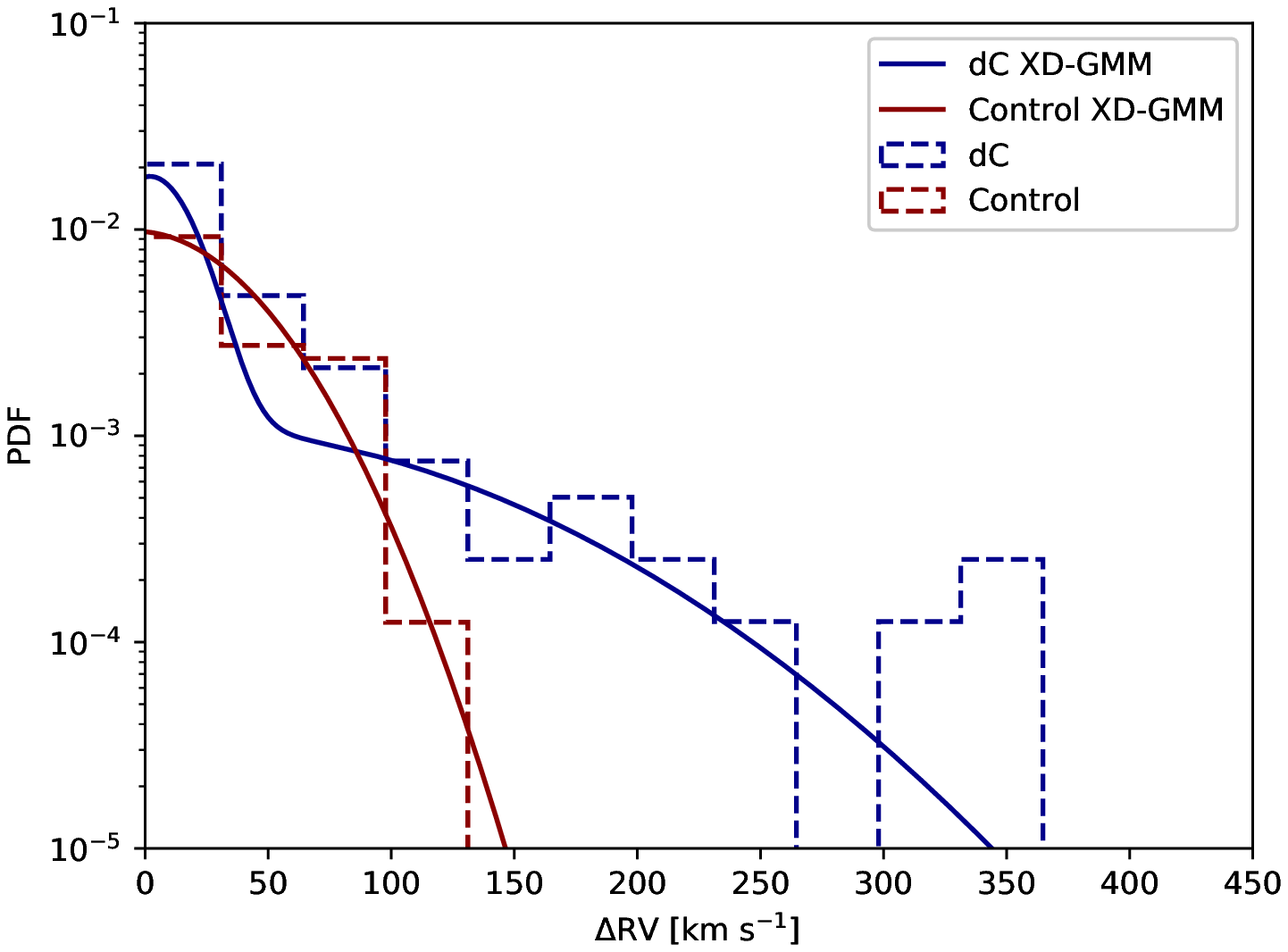}
\caption{XD-GMM for both the dC and control samples. The histograms are the measured $\Delta$RV values from this work with the dCs in  blue and the control sample in red. The smooth curves are the XD-GMM PDFs generated from the $\Delta$RV values (taking into account the $\Delta$RV errors) with the dCs in  blue and the control sample in  red (note the logarithmic scale). A central core is visible in both the dC and control samples, but the dC has an extended tail that extends from the core. This large $\Delta$RV tail is indicative of close binaries amid the dC population.}
\label{fig:XD_plot}
\end{figure}

\begin{singlespace}
\begin{deluxetable}{cDD}
\tablecaption{XD-GMM Component Fits}
\tablewidth{0.99\textwidth}
\tablehead{\colhead{Parameter} & \twocolhead{dC}& \twocolhead{Control}}
\decimals
\startdata
$\alpha_1$  &  0.688  &  1.00 \\
$\mu_1$ [km s$^{-1}$] & 2.02 &   9.69 \\
$\sigma_1$ [km s$^{-1}$] & 251.53 &  1035.82 \\
\hline
$\alpha_2$  & 0.312  &  . \\
$\mu_2$ [km s$^{-1}$] &  2.36  & . \\
$\sigma_2$ [km s$^{-1}$] & 12365.92 &  . \\
\enddata
\tablecomments{Values of the component fits for the XD-GMM for both the dC and control samples. 
Listed are the mean ($\mu$) and standard deviations ($\sigma$) of each component as well as the weights ($\alpha$; $\Sigma_i \alpha_i =1$).} 
\end{deluxetable}
\label{tab:gmm_parms}
\end{singlespace}

Figure \ref{fig:XD_plot} shows the results of the XD analysis, displaying both XD-GMMs for the two samples (smooth curves) and the histogram of the measured $\Delta$RVs (both the smooth PDFs and histograms have been normalized to an integral of one). This figure demonstrates that both the dC and control samples have a core in their $\Delta$RV distribution, but the dC distribution has a much broader wide component that flares out from the core. 

The width of the single component as fit to the control sample is wider than that narrow component of the dC sample. At the risk of over-interpreting this difference, we mention several effects that could contribute to this difference.  First, we have used the \texttt{FXCOR} reported errors, which in Section \ref{sec:xc_err} were shown to be overestimated. Since the control sample is primarily from legacy SDSS spectra with lower S/N (therefore larger errors), this overestimation is larger and may inflate the error-deconvolved core of the control distribution.  Second, the single control sample fit component must accommodate the full range of single and multiple systems.  Third, the narrow core of the dC sample could be real; perhaps, some fraction of dC binary orbits have actually widened due to processes related to mass transfer.   \citet{chen18} report that some wider binaries may undergo Bondi-Hoyle-Littleton mass transfer \citep{BHL} and further separate since orbit-synchronized rotation of the giant star could serve as an angular momentum reservoir.  However, we would still expect a narrower core for the control sample since a substantial fraction should be single stars. 

It should be noted that this XD method simply uses the GMM method with errors to determine the underlying PDF as a mixture of Gaussians, but does not imply or impose any physical meaning or model on our data.  However, it clearly allows us to determine that the dC sample has a tail that extends far beyond that of the control sample and is indicative of close binary systems included in the dC population.  

\section{Binary Orbit Simulations}
\label{sec:orbit_sim}
While our sample lacks sufficient epochs per dC to fit individual orbits, we can use the $\Delta$RV distribution to model the \added{\edit1{binary fraction and}} separation distribution.  Assuming a primitive dC mass distribution further allows us to characterize the expected period distribution of the dC sample. \replaced{We adopt a Markov Chain Monte Carlo (MCMC) simulation to fit a separation distribution to the $\Delta$RV distribution found in this work.}{\edit1{We use a Markov Chain Monte Carlo (MCMC) technique to compare these simulations to the $\Delta$RV distribution found in this work for both the dC and control samples.}}

Since little is known of dC orbital properties outside of G77-61, we make some physically-driven assumptions. First, we assume that the dC orbits have been circularized ($e = 0$) since we expect all of them to have undergone mass transfer. \replaced{Second, with circular orbits, we expect that our random epoch sampling has the same chance of catching any phase of the orbit. Therefore we sample two epochs randomly from a uniform phase distribution.}{\edit1{Second, we use the observed $\Delta$MJD distributions of each sample to simulate our observations and to sample the modeled $\Delta$RVs.}} Third, we assume that the WD mass distribution follows that found  by \citet{Kepler07} (i.e., a combination of four Gaussian components. The dominant component is centered on $0.58$ M$_{\odot}$ with a width of $0.047$ M$_{\odot}$). We use the distribution for the hot WD sample in \citet{Kepler07} since they state the distribution for the cooler WDs is not reliable. We also assume a probability density function (PDF) that is uniform over $\cos{i}$ in order to determine the PDF for $\sin{i}$. Finally, since there are no known constraints on the dC mass distribution, we assume a uniform PDF over the range of 0.2$M_{\odot}$ and 1.0$M_{\odot}$, simply assuming that dCs span the same range of masses as normal main sequence stars of the same $g-r$ color distribution. \added{\edit1{Since our control sample was selected to cover the same magnitude and color range as the dC stars, we use the same mass distribution as the dCs. The other model assumptions are also held to keep the model simplified.}}

With these assumed PDFs we simulate a population of \deleted{\edit1{dC}} stars and sample those orbits to obtain a simulated $\Delta$RV distribution. Comparing the simulated $\Delta$RV distribution to the measured one allows the MCMC to map the separation distribution parameter space. 

For the first simulation, we assumed that \replaced{our dCs have}{\edit1{our stars that are binaries have}} separations that follow a log-normal distribution with unknown mean $\mu$ and standard deviation $\sigma$, as shown in Equation \ref{eq:sep_dist}. We placed no constraints on the model parameters, aside from those required by the log-normal PDF (i.e., $\sigma \geq 0.0$ km s$^{-1}$ \added{\edit1{and $0.0 \leq f_b \leq 1.0$}}), and allowed the MCMC walkers to explore the parameter space freely.
\begin{equation}\label{eq:sep_dist}
    f(a) = \frac{1}{a\sigma \sqrt{2\pi}}\exp{\left(-\frac{(\ln{a} - \mu)^2}{2\sigma^2}\right)}
\end{equation}

\begin{figure}  
\centering
\epsscale{1.2}
\plotone{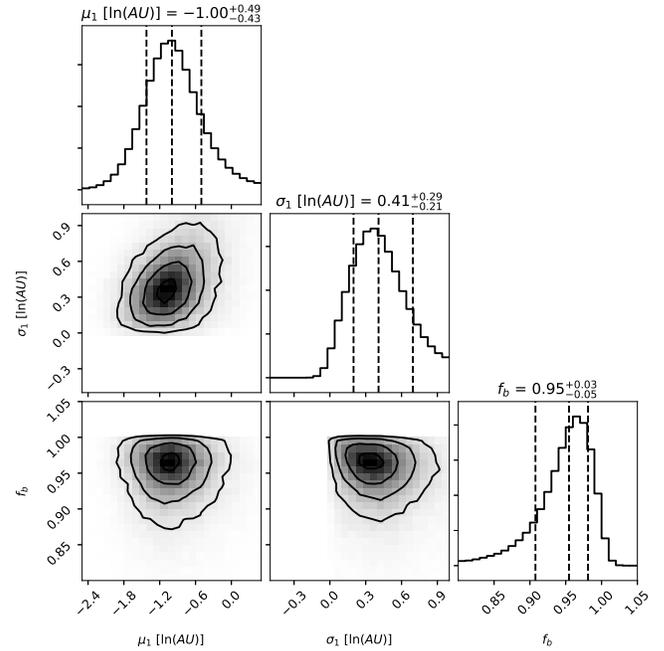}
\caption{The posterior distributions for the model parameters ($\mu, \sigma, f_b$) for the unimodal log-normal distribution from the dC sample MCMC simulation. Vertical dashed lines represent the 1$\sigma$ range and the median (50${th}$ percentile). Values are the natural logarithm ($\ln{}$) of the separation in units of AU.}
\label{fig:dC_sep_mcmc}
\end{figure}

\begin{figure}  
\centering
\epsscale{1.2}
\plotone{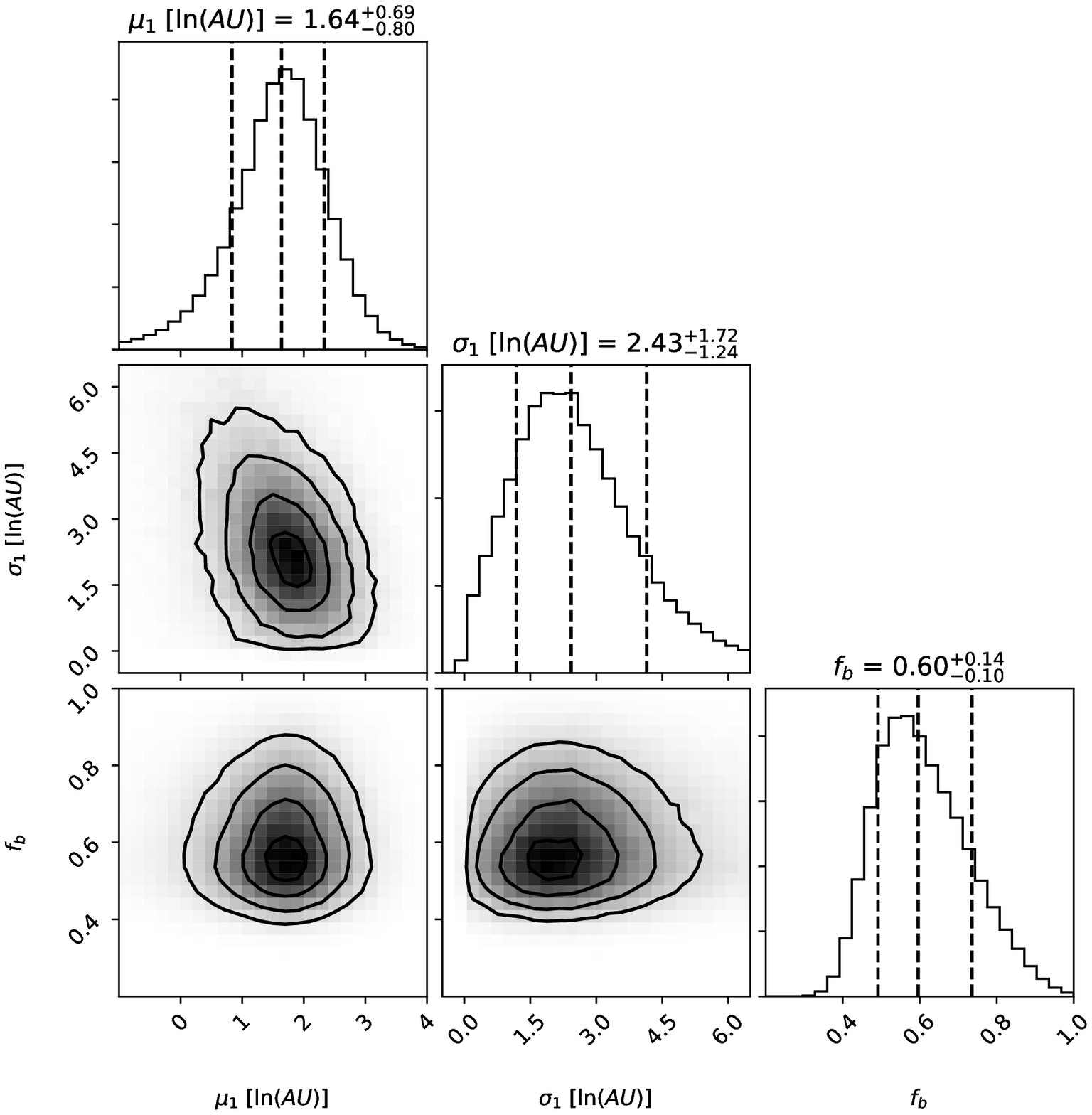}
\caption{\added{\edit1{The posterior distributions for the model parameters ($\mu, \sigma, f_b$) for the unimodal log-normal distribution from the control sample MCMC simulation. Vertical dashed lines represent the 1$\sigma$ range and the median (50${th}$ percentile). Values are the natural logarithm ($\ln{}$) of the separation in units of AU.}}}
\label{fig:control_sep_mcmc}
\end{figure}

We ran a MCMC for \replaced{100,000}{\edit1{1,000,000}} steps with 100 walkers using the \citet{gw10} algorithm. This approach allowed our walkers to explore all of the parameter space and sample the posterior of our model, which we checked for with the convergence of the chains. Figure \ref{fig:dC_sep_mcmc} shows the resulting MCMC posterior distributions for our \replaced{two}{\edit1{three}} model parameters \added{\edit1{for the dCs. Figure \ref{fig:control_sep_mcmc} shows the same plot for the control sample}}. 

\added{\edit1{From Figures \ref{fig:dC_sep_mcmc} \& \ref{fig:control_sep_mcmc}, the simulations show that the dC stars have an enhanced binary fraction as compared to the control sample (95\% vs. 60\%). The dC binary fraction fit is consistent (within the uncertainties) with a binary fraction of 100\%, indicating that dwarf carbon stars are indeed the results of binary mass transfer.}}

The resulting separation distribution from the \added{\edit1{dC}} MCMC \added{\edit1{simulation}} has a mean of 0.39 AU, a variance of 0.28 AU, and a median of 0.36 AU. These distances correspond to mean periods of 79-100 days depending on dC mass (G77-61 has a period of 245 days) and a minimum period for this distribution is on order 2.5 days (consistent with \citet{margon18}, who found a dC with a period of 2.9 days using photometry  from the Palomar Transient Factory). The separation distribution generated by our MCMC results in periods that are consistent with the \replaced{two}{\edit1{few}} periods known of individual dC systems. 

However, \citet{kool95} predicted that dC stars should follow a bimodal period distribution with one peak between $10^2 - 10^3$ days and another at $10^3 - 10^5$ days. Therefore, we also use our MCMC to model a bimodal mixture model (made of two log-normal separation distributions) of the form in Equation \ref{eq:sep_dist_bimodal}. \added{\edit1{For this model, we use our most likely dC binary fraction of 95\%.}}

\begin{multline}\label{eq:sep_dist_bimodal}
    f(a) = \frac{\alpha}{a\sigma_1 \sqrt{2\pi}}\exp{\left(-\frac{(\ln{a} - \mu_1)^2}{2\sigma_1^2}\right)} \\ + \frac{1 - \alpha}{a\sigma_2 \sqrt{2\pi}}\exp{\left(-\frac{(\ln{a} - \mu_2)^2}{2\sigma_2^2}\right)}
\end{multline}

In Equation \ref{eq:sep_dist_bimodal}, $\mu_i$ and $\sigma_i$ are the same parameters as in the unimodal distribution, and $\alpha$ is the mixing parameter in this mixture model that controls how much of each distribution contributes to the total PDF. As before, we place no constraints outside of those required by the log-normal PDFs and required by the mixing parameter (i.e. $0.0\leq{} \alpha \leq{1.0}$).

This bimodal distribution MCMC simulation was run for 100,000 steps with 100 walkers. The reduction in steps is required by increased computational load when drawing from this bimodal PDF distribution. While this change does reduce the number of points in the parameter space, the MCMC walkers still mapped the posterior quite well, which we checked for with the convergence of the chains. 

\begin{figure}  
\centering
\epsscale{1.2}
\plotone{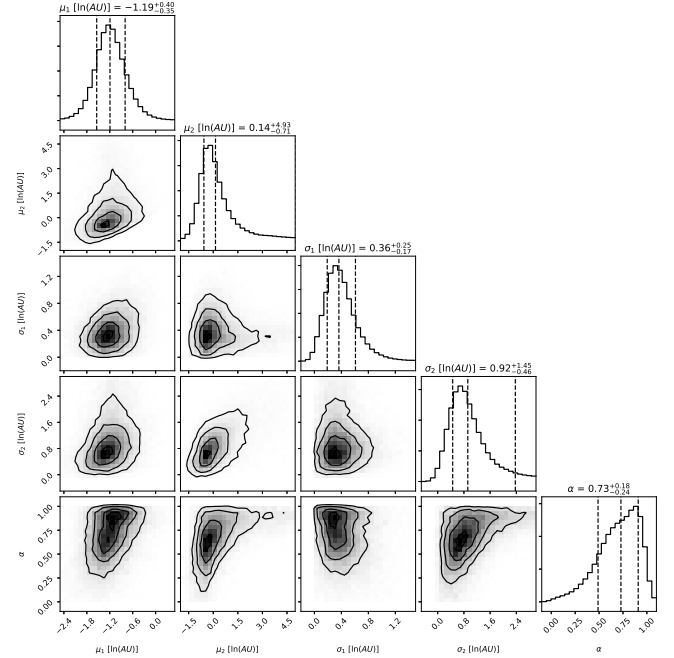}
\caption{The posterior distributions for all five of the model parameters ($\mu_i, \sigma_i, \alpha$) for the bimodal mixture model of log-normal distributions from the MCMC simulations. Vertical dashed lines represent the 1$\sigma$ range and the median (50$^{th}$ percentile). Values are the natural logarithm ($\ln{}$) of the separation in units of AU ($\alpha$ is a dimensionless mixture parameter). }
\label{fig:dC_sep_mcmc_bimod}
\end{figure}

Figure \ref{fig:dC_sep_mcmc_bimod} shows the MCMC posterior distributions for the bimodal mixture model for all five of our model parameters \added{\edit1{for the dC sample}}. In this bimodal mixture model, the total separation distribution has a mean of 0.71 AU and a variance of 1.45 AU. This distribution gives (for the previously stated uniform dC mass range) a range of the mean period of 298-413 days and a minimum period of 1.6 days. Although the number of measured dC periods is quite sparse, the period distribution (calculated from the separation distribution) is in agreement with those few periods in the literature. 

\edit1{\deleted{These results are promising, but improvements are possible. The simplest enhancement will be to use the full TDSS dC sample once all repeat spectra have been collected\deleted{\edit1{(expected in early 2019)}}. These spectra, all of high quality, will increase our sample by a factor of 3 -- 4 and should greatly improve the ability of the MCMC to determine the separation distribution (as well as provide more dCs with high $\Delta$RVs for future spectroscopy). }}

\added{\edit1{These results are promising, but improvements are possible}} One significant improvement can be achieved by measuring dC masses via orbital fits from a follow up spectroscopy campaign. Fitting an orbit to even a few dCs will place initial physical constraints on the dC mass distribution. With a more physical and realistic dC mass distribution the models and MCMC simulations can fit a more accurate separation distribution than can be done with the currently used uniform mass distribution. \edit2{\deleted{We have been awarded time on the Multiple Mirror Telescope (MMT) to begin observing our largest $\Delta$RV dCs. We have also proposed for additional MMT time to continue monitoring these dCs confirm their $\Delta$RV variability and to begin fitting orbits.}}

\section{Balmer Emission Lines}
\label{halpha}

The multi-epoch spectra present an opportunity to survey the dC sample for H$\alpha$ emission line strength and variability. Balmer line emission has been observed in dCs, and \citet{green13} found that about 2.6\% of dCs showed H$\alpha$ emission. 

There are 10 objects with H$\alpha$ emission.   Balmer line emission might be expected among dCs for several reasons: (1) coronal emission that may be a result of increased activity from spin-up during the accretion phase of the dC evolution \added{\edit2{--- valid for recent ($< 1$ Gyr) interactions before the dC has spun-down again.}}, (2) irradiation of the dC by a hot white dwarf companion, or (3) spin-orbit coupling in a close WD/dC binary. 

To explore case 1, in a related effort, we are currently analyzing {\em Chandra} observations of a small sample of dC stars to test whether their X-ray emission is consistent with dynamo rejuvenation by accretion spin-up (Green, P.J. et al. 2019, in preparation).

If the emission is from case 2, we expect to detect the WD component in the dC spectra. Indeed, all four of our DA/dCs show emission in their spectra. The remaining six of the H$\alpha$ emission line dCs are of ``normal'' type (i.e., no visible WD in the spectrum). H$\alpha$ emission is variable in only one normal dC and in none of the DA/dCs.  Since close orbits should be involved for cases 2 and 3, we will pursue further multi-epoch spectroscopy for emission line systems.


\section{Discussion}
\label{discussion}
Using multi-epoch spectroscopy we have measured the radial velocity variations of a SDSS sample of dC stars. Through MCMC methods, our modeling was able to \added{\edit1{model the binary fraction and to}} construct the separation distribution of this dC population that best recreates the observed $\Delta$RV. 

We presented the best parameters for two separation models: a unimodal log-normal distribution and a bimodal mixture model of log-normal distributions. Both models result in close binary separation distributions with means less than 1 AU, corresponding to mean periods on the order of 1 year (varying depending on dC mass). 

Our sample contains a handful of objects with large ($\geq$ 100 km s$^{-1}$) $\Delta$RV measurements that are indicative of close binary systems. These objects will be targeted for future spectroscopy to constrain orbital parameters thereby better characterizing the separation distribution. In addition, orbital fits will also allow us to determine the masses of the dCs assuming a WD component.

\citet{badenes18} analyze the RV variability of main sequence stars and report that the binary fraction is likely higher for more metal-poor stars.  Carbon stars are suspected to form more easily at lower metallicity; indeed, about 20\% of stars with \replaced{[Fe/H]$<−$2}{\edit1{[Fe/H]$< -2$}} show carbon-enhancement (e.g., \citealt{christlieb01, lee13}), but that frequency is increasing rapidly as metallicity decreases \citep{placco14}. Close binaries ($<10$ AU) also show increases in lower metallicity populations \citep{moe18}. The dC in G77-61 is thought to be extremely metal-poor \citep{gass88}.   The measured dC $\Delta$RV distribution being wider than the control sample could in part be due both to low metallicity and to evolutionary effects since dC stars are carbon-enhanced by binary mass transfer.  If the mass transfer results in inward evolution of the binary, then that should further widen the $\Delta$RV distribution for dCs.  Binaries with an AGB primary can be at large separations and still, via wind-Roche lobe overflow, lose orbital angular momentum, evolve towards direct Roche lobe overflow and/or tidal friction towards a common envelope \citep{chen18}.  Therefore, the fraction of binaries that result in mass transfer in a common envelope and a tight binary configuration may be quite large. The dC stars present a population of post mass transfer binaries that are unusually easy to identify, but may represent just a tiny fraction of such stars --- those sufficiently cool and with large enough C/O to produce C$_2$ and/or CN bands. In some cases, the AGB evolution may have been truncated during the common envelope phase before significant carbon dredge up.  A much larger space density of post mass transfer M dwarfs may remain unidentified until massive multi-epoch RV surveys become available.  

\facility{Sloan 2.5-meter}

\software{\texttt{Astropy} \citep{astropy}, \texttt{corner} \citep{corner}, \texttt{IRAF} \citep{iraf}, \texttt{matplotlib} \citep{matplotlib}, \texttt{Numpy} \citep{numpy}, \texttt{Scipy} \citep{scipy}, \texttt{Scikit-learn} \citep{Scikit-learn}}

\acknowledgments
We thank John Ruan for longstanding TDSS contributions, important aspects for this research.

Funding for the Sloan Digital Sky Survey IV has been provided by the
Alfred P. Sloan Foundation, the U.S. Department of Energy Office of
Science, and the Participating Institutions. SDSS acknowledges
support and resources from the Center for High-Performance Computing at
the University of Utah. The SDSS web site is www.sdss.org.

SDSS is managed by the Astrophysical Research Consortium for the Participating Institutions of the SDSS Collaboration including the Brazilian Participation Group, the Carnegie Institution for Science, Carnegie Mellon University, the Chilean Participation Group, the French Participation Group, Harvard-Smithsonian Center for Astrophysics, Instituto de Astrof\'{i}sica de Canarias, The Johns Hopkins University, Kavli Institute for the Physics and Mathematics of the Universe (IPMU) / University of Tokyo, Lawrence Berkeley National Laboratory, Leibniz Institut f\"{u}r Astrophysik Potsdam (AIP), Max-Planck-Institut f\"{u}r Astronomie (MPIA Heidelberg), Max-Planck-Institut f\"{u}r Astrophysik (MPA Garching), Max-Planck-Institut f\"{u}r Extraterrestrische Physik (MPE), National Astronomical Observatories of China, New Mexico State University, New York University, University of Notre Dame, Observat\'{o}rio Nacional / MCTI, The Ohio State University, Pennsylvania State University, Shanghai Astronomical Observatory, United Kingdom Participation Group, Universidad Nacional Aut\'{o}noma de M\'{e}xico, University of Arizona, University of Colorado Boulder, University of Oxford, University of Portsmouth, University of Utah, University of Virginia, University of Washington, University of Wisconsin, Vanderbilt University, and Yale University.

\bibliographystyle{yahapj}
\bibliography{main}

\listofchanges
\end{document}